\documentclass[]{spie}  

\usepackage{amsmath,amsfonts,amssymb}
\usepackage{graphicx}
\usepackage[colorlinks=true, allcolors=blue]{hyperref}
\usepackage[italian,english]{babel} 
\usepackage{epstopdf}

\newcommand{\degrees} {$^\circ$}
\newcommand{\arcsec}{$^{\prime\prime}$}

\def\aj{AJ}                   
\def\apj{ApJ}                 
\def\apjl{ApJ}                
\def\apjs{ApJS}               
\def\aap{A\&A}                
\def\mnras{MNRAS}             

\def\pasj{PASJ}               
\def\procspie{Proc.~SPIE}     

\title{Prospects of Deep Field Surveys with Global-MCAO on an ELT}
\author[*a,b]{Elisa Portaluri}
\author[a,b]{Valentina Viotto}
\author[a,b]{Roberto Ragazzoni}
\author[a,b]{Marco Gullieuszik}
\author[a,b]{Maria Bergomi}
\author[a,b]{Federico Biondi}
\author[a,b]{Elena Carolo}
\author[a,b]{Simonetta Chinellato}
\author[a,b]{Marco Dima}
\author[a,b]{Jacopo Farinato}
\author[a,b]{Davide Greggio}
\author[a,b]{Demetrio Magrin}
\author[a,b]{Luca Marafatto}
\author[a,b,c]{Gabriele Umbriaco}
\author[a,b,c]{Daniele Vassallo}
\affil[a]{INAF - Osservatorio Astronomico di Padova, vicolo dell'Osservatorio 5, Padova, Italy}
\affil[b]{ADONI - Laboratorio Nazionale Ottiche Adattive, Italy}
\affil[c]{Dipartimento di Fisica e Astronomia, Universit\`a degli Studi di Padova, Italy}
\authorinfo{*elisa.portaluri@oapd.inaf.it}

\begin{document} 
\maketitle

\begin{abstract}
Several astronomical surveys aimed at the investigation of the extragalactic components were carried out in order to map systematically the universe and its constituents. An excellent level of detail is needed, and it is possible only using space telescopes or with the application of adaptive optics (AO) techniques for ground-based observatories.
By simulating $K$-band observations of 6000 high-redshift galaxies in the {\it Chandra Deep Field South} region, we have already shown how an extremely large telescope can carry out photometric surveys successfully using the Global-MCAO, a natural guide stars based technique that allows the development of extragalactic research, otherwise impracticable without using laser guide stars.
As the outcome of the analysis represents an impact science case for the new instruments on upcoming ground-based telescopes, here we show how the investigation of other observed deep fields could profit from such a technique. Further to an overview of the surveys suitable for the proposed approach, we show preliminary estimations both on geometrical (FoV and height) and purely AO perspectives (richness and homogeneity of guide stars in the area) for planned giant telescopes. 
\end{abstract}


\section{INTRODUCTION}
\label{sec:intro}
Understanding the formation and evolution of galaxies is a lively debated topic in astrophysics, because the advent of more powerful telescopes stimulated more detailed theories and numerical simulations to interpret new observations. The main issue concerns the role of the competing scenarios in shaping galaxies: large protogalaxies formed through a dissipational collapse according to the monolithic scenario, whereas galaxies are the result of successive merging between small structures in the hierarchical merging process. Cold or hot-flow accretion, mergers involving gas or not (dissipational or dissipationless), feedback and outflows, and rapid monolithic aggregation are paradigms that live together within theoretical frameworks for understanding the early formation, structural properties, and evolution of galaxies. Observations spanning the full history of the Universe are critical for developing the knowledge of these processes: the early building blocks of galaxies can be studied at {\it z} $>6$, earlier redshift galaxies ($1<z<4$) span the peak of the massive galaxy building era, while, locally, dwarf galaxies can provide a fossil bed of relics from the low-mass end of the galaxy formation spectrum.

\begin{table}[!htbp]
\caption{Observing parameters for some of the most-studied surveys. Column~1: Name of the survey. Column~2: Right ascension and Column~3: Declination of the central pointing.  Column~4: Field of view area.  Column~5: Reference. } 
\label{tab:surveys}
\begin{center}       
\begin{tabular}{|c|c|c|c|c|} 
\hline
\rule[-1ex]{0pt}{3.5ex} Survey name  & RA          &  DEC        & Field of View & Reference \\
\rule[-1ex]{0pt}{3.5ex}              & [$^{\circ}$] & [$^{\circ}$] & [square arcmin]  &           \\
\hline	
\rule[-1ex]{0pt}{3.5ex} HDF	    &  189.2042   &  62.2161   &       5   &  \citenum{Williams1996} \\	
\hline	
\rule[-1ex]{0pt}{3.5ex} NDWFS-Bootes &  217.500   &  34.5000   &   3.24$10^4$   &   \citenum{Jannuzi1999} \\	
\hline	
\rule[-1ex]{0pt}{3.5ex} NDWFS-Cetus &   31.8708   &  -4.7356   &   3.31$10^4$   &   \citenum{Jannuzi1999} \\	
\hline	
\rule[-1ex]{0pt}{3.5ex} HDF-S       &  338.2458   & -60.5508   &       5.3 &  \citenum{Williams2000} \\	
\hline	
\rule[-1ex]{0pt}{3.5ex} CDF-S	    &   53.1167   & -27.8083   &     3.96$10^2$   &   \citenum{Giacconi2001} \\	
\hline	
\rule[-1ex]{0pt}{3.5ex} DLS-1       &   13.3542   &  12.5653   &   1.44$10^4$   &   \citenum{Wittman2002} \\	
\hline	
\rule[-1ex]{0pt}{3.5ex} DLS-2       &  139.5000   &  30.0000   &   1.44$10^4$   &   \citenum{Wittman2002} \\	
\hline	
\rule[-1ex]{0pt}{3.5ex} DLS-3       &   80.0000   & -49.0000   &   1.44$10^4$   &   \citenum{Wittman2002} \\	
\hline	
\rule[-1ex]{0pt}{3.5ex} DLS-4       &  163.0000   &  -5.0000   &   1.44$10^4$   &   \citenum{Wittman2002} \\	
\hline	
\rule[-1ex]{0pt}{3.5ex} DLS-5       &  208.7500   & -10.0000   &   1.44$10^4$   &   \citenum{Wittman2002} \\	
\hline	
\rule[-1ex]{0pt}{3.5ex} DLS-6       &   32.5000   &  -4.5000   &   1.44$10^4$   &   \citenum{Wittman2002} \\	
\hline	
\rule[-1ex]{0pt}{3.5ex} DLS-7       &  218.0000   &  34.2800   &   1.44$10^4$   &   \citenum{Wittman2002} \\	
\hline	
\rule[-1ex]{0pt}{3.5ex} EGSS	    &  214.2500   &  52.5000   &     7.00$10^2$   &   \citenum{Davis2003} \\	
\hline	
\rule[-1ex]{0pt}{3.5ex} GEMS        &   53.1042   & -27.8139   &     8.00$10^2$  &   \citenum{Rix2004} \\
\hline	
\rule[-1ex]{0pt}{3.5ex} GOODS-N     &  189.2292   &  62.2375   &     1.60$10^2$   &   \citenum{Giavalisco2004} \\	
\hline	
\rule[-1ex]{0pt}{3.5ex} GOODS-S     &   53.1250   & -27.8056   &     1.60$10^2$   &   \citenum{Giavalisco2004} \\	
\hline	
\rule[-1ex]{0pt}{3.5ex} SubaruDF    &  201.1625   &  27.4906   &     9.18$10^2$   &   \citenum{Kashikawa2004} \\	
\hline	
\rule[-1ex]{0pt}{3.5ex} COSMOS      &  150.1167   &   2.2058   &    7.200$10^3$   &   \citenum{Scoville2004} \\	
\hline
\rule[-1ex]{0pt}{3.5ex} GSS	    &  214.4042   &  52.4828   &     1.27$10^2$   &   \citenum{Vogt2005} \\	
\hline	
\rule[-1ex]{0pt}{3.5ex} HUDF        &   53.1625   & -27.7914   &      11   &   \citenum{Beckwith2006} \\	
\hline	
\rule[-1ex]{0pt}{3.5ex} CFHTLS-D1   &   36.4958   &  -4.4944   &    3.60$10^3$   &   \citenum{Cuillandre2006} \\	
\hline	
\rule[-1ex]{0pt}{3.5ex} CFHTLS-D2   &  150.1167   &   2.2083   &    3.60$10^3$   &   \citenum{Cuillandre2006} \\	
\hline	
\rule[-1ex]{0pt}{3.5ex} CFHTLS-D3   &  214.8625   &  52.6822   &    3.60$10^3$   &   \citenum{Cuillandre2006} \\	
\hline	
\rule[-1ex]{0pt}{3.5ex} CFHTLS-D4   &  333.8792   & -17.7320   &    3.60$10^3$   &   \citenum{Cuillandre2006} \\	
\hline	
\rule[-1ex]{0pt}{3.5ex} AEGIS	    &  214.2500   &  52.5000   &     7.00$10^2$   &   \citenum{Davis2007} \\	
\hline	
\rule[-1ex]{0pt}{3.5ex} UKIDSS-DXS1 &   36.2500   &  -4.5000   &  1.26$10^5$   &   \citenum{Lawrence2007} \\	
\hline	
\rule[-1ex]{0pt}{3.5ex} UKIDSS-DXS2 &  164.2500   &  57.6667   &  1.26$10^5$   &   \citenum{Lawrence2007} \\	
\hline	
\rule[-1ex]{0pt}{3.5ex} UKIDSS-DXS3 &  242.5      &  54.0000   &  1.26$10^5$   &   \citenum{Lawrence2007} \\	
\hline	
\rule[-1ex]{0pt}{3.5ex} UKIDSS-DXS4 &  334.2500   &   0.3333   &  1.26$10^5$   &   \citenum{Lawrence2007} \\	
\hline	
\rule[-1ex]{0pt}{3.5ex} UKIDSS-UD   &   36.2500   &  -4.5000   &    2.77$10^3$   &   \citenum{Lawrence2007} \\	
\hline	
\rule[-1ex]{0pt}{3.5ex} HUDF9       &   53.1625   & -27.7914   &       4.7 &   \citenum{Bouwens2011} \\	
\hline	
\rule[-1ex]{0pt}{3.5ex} C\_GOODS-N   &  189.2286  &  62.2385   &     9.18$10^2$   &   \citenum{Koekemoer2011} \\	 
\hline	
\rule[-1ex]{0pt}{3.5ex} C\_GOODS-S   &   53.1228  & -27.8050   &    1.12$10^3$   &   \citenum{Koekemoer2011} \\	
\hline	
\rule[-1ex]{0pt}{3.5ex} C\_COSMOS    &  150.1163  &   2.2010   &   1.44$10^4$   &   \citenum{Koekemoer2011} \\	
\hline	
\rule[-1ex]{0pt}{3.5ex} C\_EGS       &  214.8250  &  52.8250   &    7.20$10^3$   &   \citenum{Koekemoer2011} \\	
\hline	
\rule[-1ex]{0pt}{3.5ex} C\_UDS       &   34.4062 &  -5.2000    &    7.20$10^3$   &   \citenum{Koekemoer2011} \\
\hline	
\rule[-1ex]{0pt}{3.5ex} UDF12       &   53.1625   & -27.7914   &       5   &   \citenum{Koekemoer2013} \\	
\hline	
\rule[-1ex]{0pt}{3.5ex} XDF         &   53.1583   & -27.7833   &      10.9 &   \citenum{Illingworth2013} \\
\hline	
\end{tabular}
\end{center}
\end{table} 
Deep field surveys answered to several astrophysical questions, providing a wide database of objects at large and intermediate redshifts.  In fact, in about 20 years of observations, a lot of sky regions were mapped and Table~\ref{tab:surveys} shows a list of them that can not be exhaustive. Some of these runs were combined together in order to merge the information in the different bands, making a synergy that is fundamental in order to have a comprehensive view of Universe.
In this framework, the NASA's Great Observatories Program can be considered one of the best example of this synergy, being a mission designed to examine specific wavelength/energy regions of the electromagnetic spectrum using different technologies and observatories.

A number of future space- and ground-based facilities will definitely contribute to this topic, starting from the next generation of the ELTs, such as the European Extremely Large Telescope \cite{Gilmozzi2007}, the Giant Magellan Telescope \cite{Johns2008} and the Thirty Meter Telescope \cite{Szeto2008}, whose main characteristics are listed in Table~\ref{tab:telescopes}.

Moreover, in order to widely progress in this field and to improve the potentialities, two routes can be followed: the sole improvement of the engineering side of the existing instruments, or the attempt to turn into reality novel concepts and new ideas. Being the latter an exciting challenge, [\citenum{Ragazzoni2010}] proposed a new Adaptive Optics technique, called Global-Multi Conjugate Adaptive Optics (GMCAO), in which a wide field of view (labelled technical FoV) can be used to look for natural guide stars and correct the smaller scientific FoV. For a review of the concept and the implementation of an example of such a system, see [\citenum{Viotto2015}].

\begin{table}[!htbp]
\caption{Observing capabilities of the ELTs. Column~1: Telescope Name. Column~2: Diameter of the primary mirror. Column~3: Site with coordinates.  Column~4: Telescope field of view.  Column~5: Resolution. Column~6: Filters.} 
\label{tab:telescopes}
\begin{center}       
\begin{tabular}{|c|c|c|c|c|c|} 
\hline
\rule[-1ex]{0pt}{3.5ex} Telescope             & Diameter      & Site                           & Field of View                    & Resolution & Filters\\
\rule[-1ex]{0pt}{3.5ex}                       & [m]           & [$^{\circ}$]                    &                                  & [mas]      &        \\
\hline	
\rule[-1ex]{0pt}{3.5ex} E-ELT (MICADO)        & 39            & Cerro Armazones (-70.19,-24.59) &  53\arcsec $\times$ 53\arcsec    & 4   & I z Y J H K  \\
\rule[-1ex]{0pt}{3.5ex}    	              &               &                                &  16\arcsec $\times$ 16\arcsec    & 1.5 &              \\	
\hline	
\rule[-1ex]{0pt}{3.5ex} GMT (GMTIFS)	      & 25.4          & Las Campanas (-70.69,-29.02)   &  20.4\arcsec$\times$ 20.4\arcsec & 5   & J H K        \\	
\hline
\rule[-1ex]{0pt}{3.5ex} TMT (IRIS)	      & 30       & Mauna Kea (-155.47,19.82)           &  34\arcsec $\times$ 34\arcsec    & 4   & J H K        \\
                                              &          & La Palma (-17.89,28.89)             &                                  &     &              \\	
\hline
\end{tabular}
\end{center}
\end{table}

[\citenum{Portaluri2017}] presented a scientific case in the area of applicability of the GMCAO technique, recovering the structural parameters of a sample of 6000 synthetic high-redshift galaxies observed in the {\it Chandra Deep Field South} region with a GMCAO-assisted extremely large telescope (ELT) and performing the source detection and two-dimensional fitting analysis. These studies have shown that the GMCAO approach can produce robust results when studying the photometry of extragalactic fields and can provide a useful frame of reference for a number of science cases. Here, we want to investigate if it is possible to extend the same evaluation to other deep fields, and therefore if GMCAO can be used to map all the regions of the sky. The selection of some representative surveys and their observabilities with the planned ELTs are shown in Section~\ref{sec:obs}, while in Section~\ref{sec:stars} preliminary estimations on AO perspectives are given.

\section{Surveys selection and objects visibility}
\label{sec:obs}
\begin{figure} [!htbp]
   \begin{center}
     \begin{tabular}{c}
     \includegraphics[width=14cm]{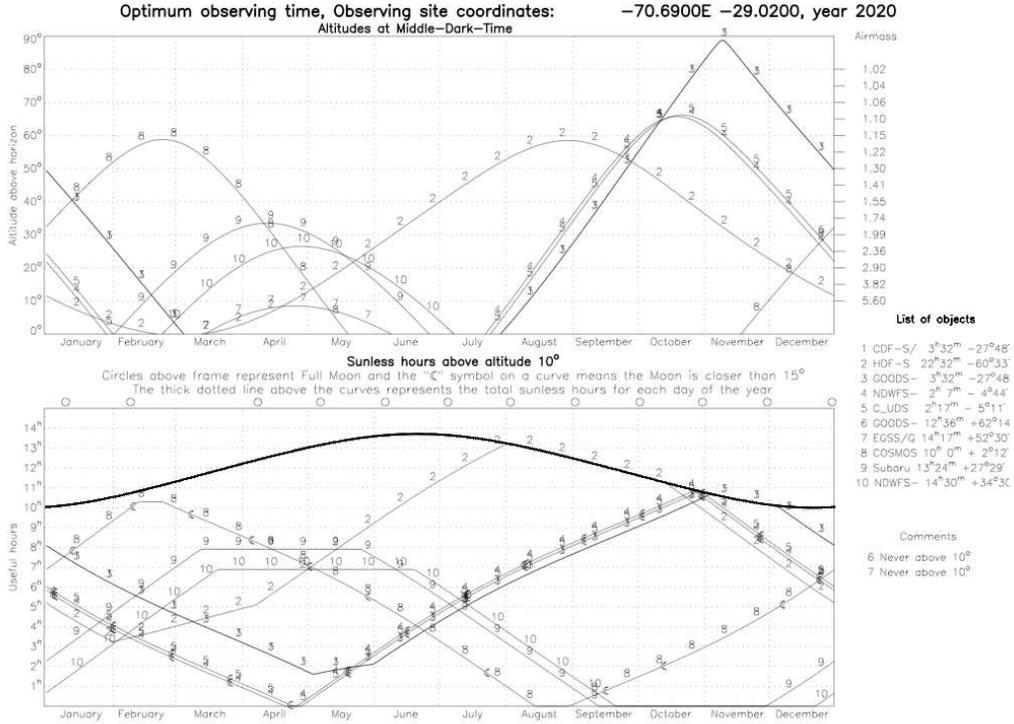}
   \end{tabular}
   \end{center}
   \caption{Objects visibility for the 10 selected surveys from the GMT site coordinates. Image courtesy of the Isaac Newton Group of Telescopes, La Palma.} 
  {\label{fig:starobs1}}
\end{figure}

\begin{figure} [!htbp]
   \begin{center}
     \begin{tabular}{c}
     \includegraphics[width=14cm]{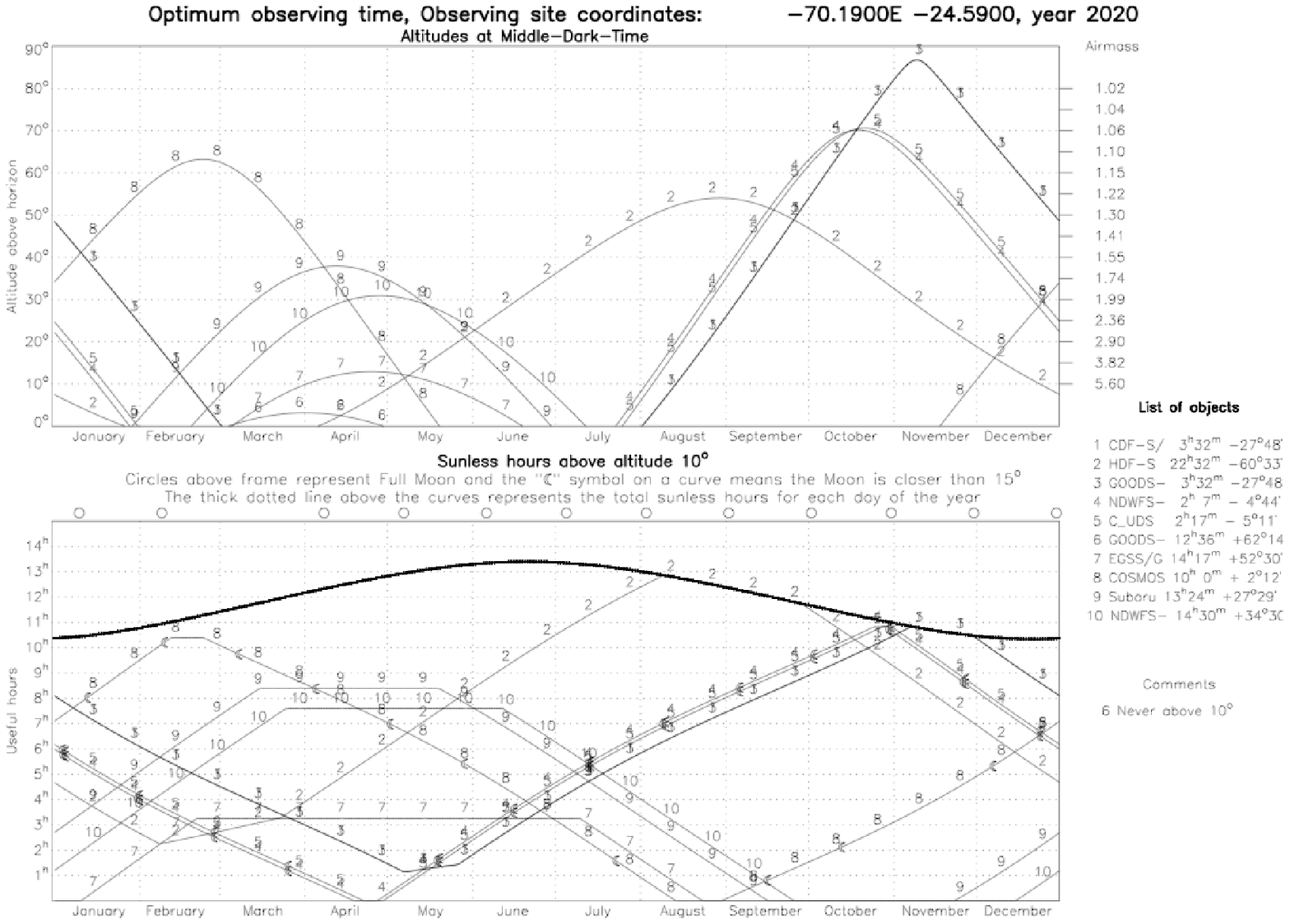}
   \end{tabular}
   \end{center}
   \caption{As in Figure~\ref{fig:starobs1} but referred to the ELT site coordinates. Image courtesy of the Isaac Newton Group of Telescopes, La Palma.} 
  {\label{fig:starobs2}}
\end{figure}

\begin{figure} [!htbp]
   \begin{center}
     \begin{tabular}{c}
     \includegraphics[width=14cm]{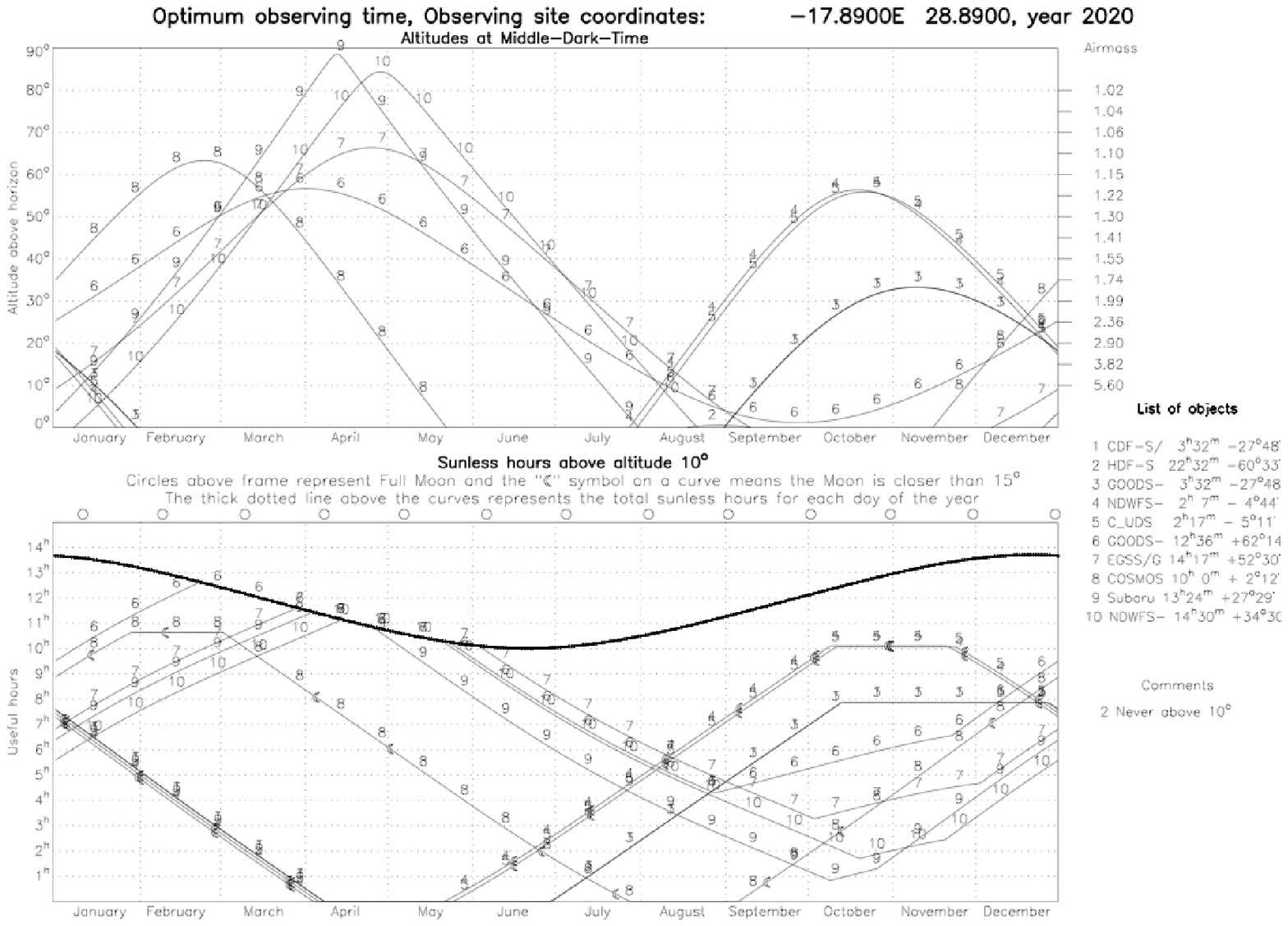}
   \end{tabular}
   \end{center}
   \caption{As in Figure~\ref{fig:starobs1} but referred to the TMT site coordinates. Image courtesy of the Isaac Newton Group of Telescopes, La Palma.} 
  {\label{fig:starobs3}}
\end{figure}

To have a hint of the visibility of the most-studied deep fields with GMCAO, we selected 10 surveys in 10 different regions of the sky: CDF-S (already studied by [\citenum{Portaluri2017}]), HDF-S, GOODS-S, NDFS-Cetus, C\_UDF,GOODS-N, EGSS/CFHTLS-D3, COSMOS/CFHTLS-D2, SubaruDF, and NDWFS-Bootes. We used the tool STAROBS\footnote{http://catserver.ing.iac.es/staralt/} that measures the observability of objects plotting how altitude changes over a year according to the ELTs site coordinates, as shown in Figure~\ref{fig:starobs1},~\ref{fig:starobs2},~\ref{fig:starobs3}.
Depending on the mutual position (site and sky coordinates), the surveys we have chosen can be considered as perfectly visible, even if we exclude altitudes below 30\degrees, for instrumental limits and for the high value of the airmass.


\section{Geometrical and AO consideration}
\label{sec:stars}
\begin{figure} [!htbp]
   \begin{center}
     \begin{tabular}{c}
       \includegraphics[width=8cm]{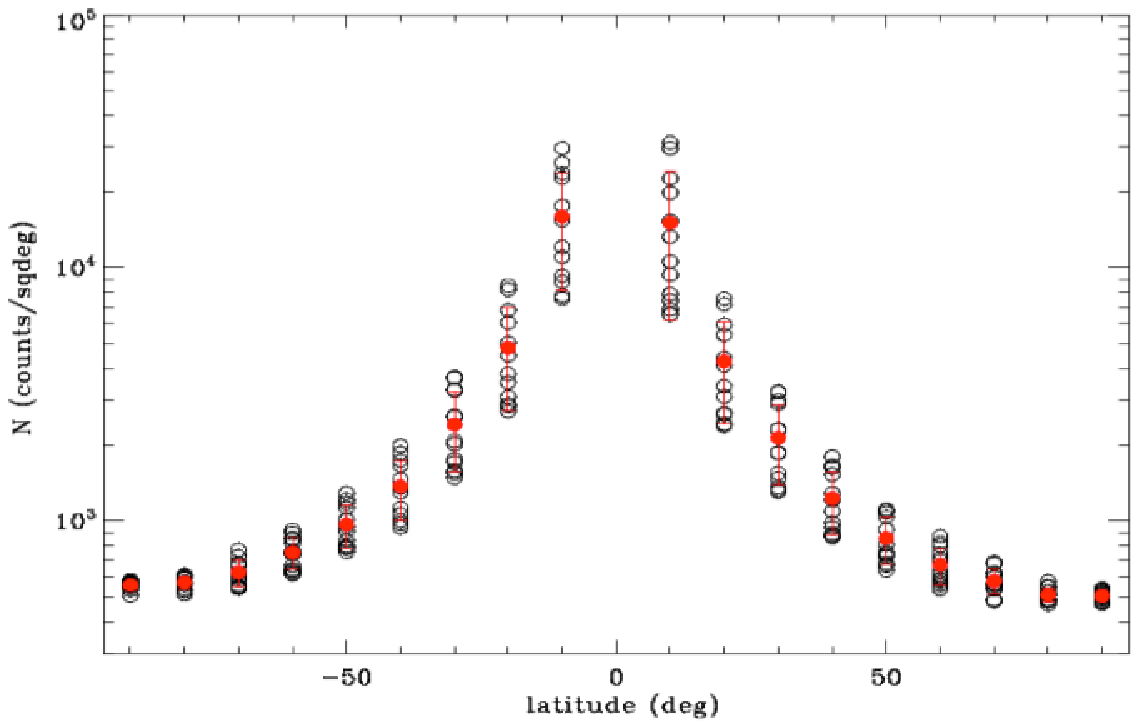}
       \includegraphics[width=8cm]{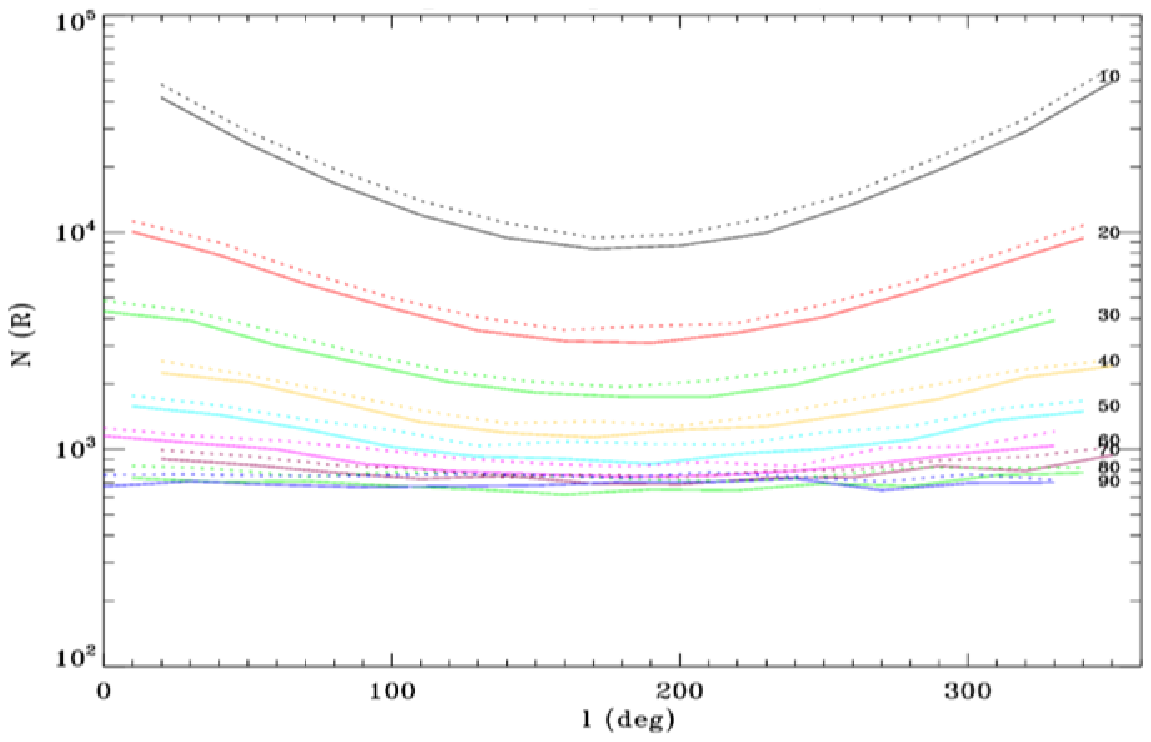}
   \end{tabular}
   \end{center}
   \caption{Left-hand panel: star density per square degree as a function of the latitude. The scattering is due to different positions over the grid. The red points represent the mean. Right-hand panel: star density per square degree as a function of the longitude. Dashed lines are for the southern hemisphere.} 
  {\label{fig:trilegal}}
\end{figure}

\begin{figure} [!htbp]
   \begin{center}
     \begin{tabular}{c}
       \includegraphics[width=8cm]{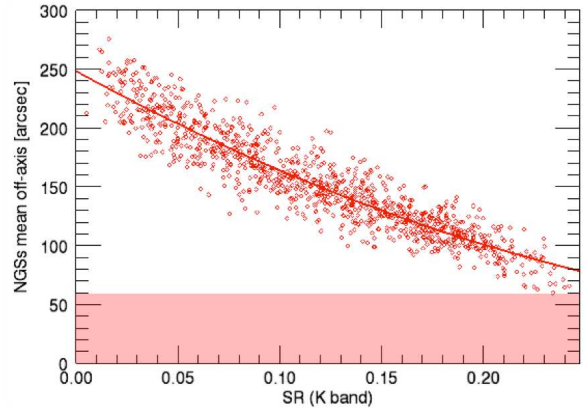}
   \end{tabular}
   \end{center}
   \caption{SR that can be obtained with GMCAO according to the asterism maximum radius of the natural guide stars.} 
  {\label{fig:asterism}}
\end{figure}

In our simulations, the GMCAO method is implemented and tested using 6 natural guide stars, therefore the knowledge of the stellar density over the sky is a fundamental request for the applicability of such a technique. First, we decided to make a theoretical simulation using TRILEGAL~\cite{Girardi2005}, a population synthesis code that simulates the stellar photometry of our Galaxy: we built a grid in the whole sky and recover the number of stars within 1 square degree with $R > 18$ mag.
Figure~\ref{fig:trilegal} shows the dependence of the star density according to the galactic latitude and longitude. As expected, the latitude plays a key role, however in all the cases, there will be enough stars to apply GMCAO. Obviously, the more stars available will be, the better performance we will expect because the possibility to select a good asterism is higher, as reported in Figure~\ref{fig:asterism}, which also shows that, beyond the increase of the encircled energy, a SR of 30\% can be reached. The region above 60 arcsec is excluded because we did not consider stars within the scientific field and also we rejected stars that are close each other less than 10 arcsec to avoid overlapping effects.


\begin{figure} [!htbp]
   \begin{center}
     \begin{tabular}{c}
       \includegraphics[width=12cm]{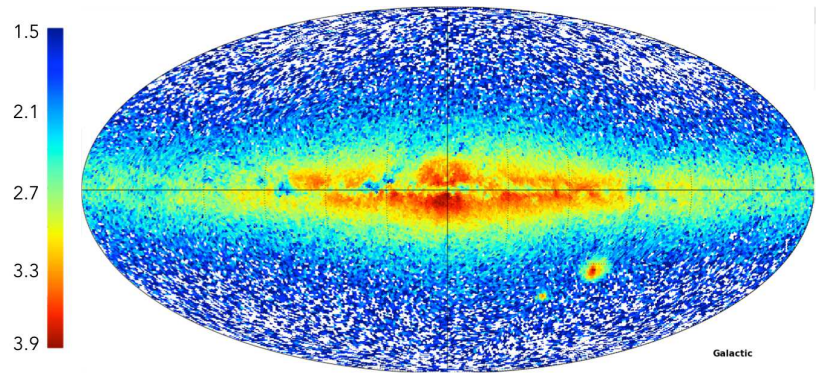}
   \end{tabular}
   \end{center}
   \caption{Logarithm of the GAIA DR1 density of stars with $8<g<18$ within the GMCAO technical field area over the whole sky.} 
  {\label{fig:gaia}}
\end{figure}

Moreover, we can use the GAIA first release (DR1, [\citenum{Brown2016}]) to access real data of stars with {\it g} magnitude between 8 and 18 and measure at random positions the star density within the technical FoV area (78.7 square arcmin), as shown in Figure~\ref{fig:gaia}. White regions are areas not covered by our random investigation.
As expected, the Milky Way plane and the Magellanic Clouds have the maximum concentration of stars, confirming the theoretical predictions done using TRILEGAL.

 \section{Conclusions}
 \label{sec:concl}
The era of the next generation of giant telescopes requires not only the advent of new technologies but also the development of novel methods, in order to exploit fully the extraordinary potential they are built for. GMCAO pursues this approach, with the goal of achieving good performance over a field of view of a few arcmin and an increase in sky coverage.
In this work we have shown that GMCAO is a reliable approach to assist ELT observations of extragalactic interest, especially for the photometric survey strategy.
This technique can be applicable to all the sky, as theoretical counts and observational data show, giving a gain in terms of increase of the encircled energy and SRs up to 30\% for the best asterisms.


\end{document}